\documentclass{ecai} 

\usepackage{latexsym}
\usepackage{amssymb}
\usepackage{amsmath}
\usepackage{amsthm}
\usepackage{booktabs}
\usepackage{enumitem}
\usepackage{graphicx}
\usepackage{color, xcolor}

\usepackage{array}

\usepackage{multirow}
\usepackage{makecell}
\usepackage{algorithm}
\usepackage{algorithmic}
\usepackage{booktabs}
\usepackage{inputenc}

\newcommand{\BibTeX}{B\kern-.05em{\sc i\kern-.025em b}\kern-.08em\TeX}
\newcommand{\minor}[1]{\color{black}#1}

\begin{document}


\begin{frontmatter}


\paperid{130} 


\title{Search, Examine and Early-Termination: Fake News Detection with Annotation-Free Evidences}

\author[A]{\fnms{Yuzhou}~\snm{Yang}}
\author[A]{\fnms{Yangming}~\snm{Zhou}}
\author[A,B]{\fnms{Qichao}~\snm{Ying}} 
\author[A]{\fnms{Zhenxing}~\snm{Qian}\thanks{Corresponding Author. Email: zxqian@fudan.edu.cn.}}
\author[A]{\fnms{Xinpeng}~\snm{Zhang}} 

\address[A]{School of Computer Science, Fudan University}
\address[B]{NVIDIA}


\begin{abstract}
Pioneer researches recognize evidences as crucial elements in fake news detection apart from patterns. Existing evidence-aware methods either require laborious pre-processing procedures to assure relevant and high-quality evidence data, or incorporate the entire spectrum of available evidences in all news cases, regardless of the quality and quantity of the retrieved data. In this paper, we propose an approach named \textbf{SEE} that 
\minor{retrieves useful information from web-searched annotation-free evidences}
with an early-termination mechanism. The proposed SEE is constructed by three main phases: \textbf{S}earching online materials using the news as a query and directly using their titles as evidences without any annotating or filtering procedure, sequentially \textbf{E}xamining the news alongside with each piece of evidence via attention mechanisms to produce new hidden states with \minor{retrieved} information, and allowing \textbf{E}arly-termination within the examining loop by assessing whether there is adequate confidence for producing a correct prediction. We have conducted extensive experiments on datasets with unprocessed evidences, i.e., Weibo21, GossipCop, and pre-processed evidences, namely Snopes and PolitiFact. The experimental results demonstrate that the proposed method outperforms state-of-the-art approaches.
\end{abstract}

\end{frontmatter}


\section{Introduction}
The ubiquitous availability and accessibility of the Internet have reshaped the way people obtain and engage with information. Yet, it has concurrently paved the way for the rapid propagation of fake news, which can swiftly amass momentum on social media and various online platforms. 
The propagation of false information has the potential to yield ill-informed perspectives, with serious implications including public opinion manipulation, concealing the truth, and the incitement of crimes.

\begin{figure}[!t]
  \centering
  \includegraphics[width=1.0\linewidth]{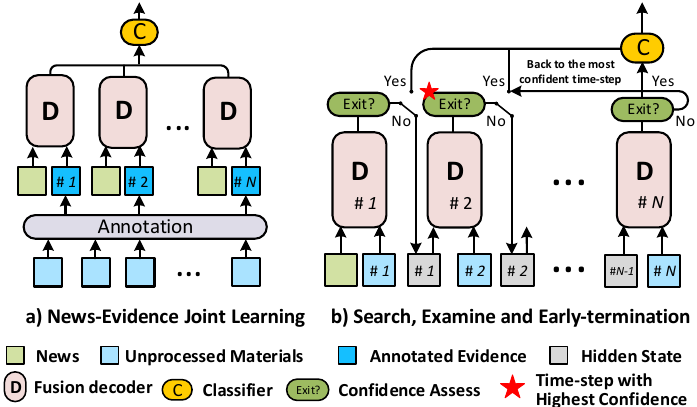}
  \caption{Comparison of methodology between a) previous news-evidence joint learning schemes, \minor{typically requiring human-participated evidences and utilizing them altogether}, and b) the proposed ``Search, Examine and Early-termination'' (SEE) scheme, \minor{which} is capable of utilizing \minor{web}-searched materials as evidences without annonating and sequentially examining the news alongside each piece of evidence with Early-termination. \# denotes the serial number. } 
  \label{fig:intro}
\end{figure}
Fake News Detection (FND) involves analyzing the probability of news containing misconducting information.
Early methods mainly utilize low-level coarse statistical analyses of news content to estimate the veracity of news,
e.g., punctuation, lexical statistics~\cite{castillo2011information} or grammar~\cite{feng2012syntactic}.
With the evolution of machine learning techniques, hand-crafted pattern analyzers have been largely supplanted by deep networks.
The prevailing methodology is to prepare thousands of real and fake news samples, extracting features from each sample, and constructing a classifier to establish a binary classification boundary.
The literature has had success stories from either mining linguistic features, e.g., pragmatic pattern~\cite{chen2015misleading}, writing style~\cite{yu2017convolutional}, sentiment~\cite{zhang2021mining}, or visual features, e.g., image quality~\cite{jin2016novel}, forgery artifacts~\cite{Leveraging_5}, joint spatial-temporal features~\cite{qi2019exploiting}.
Nevertheless, these methods still rely exclusively on pattern analysis of static news, where the learned abnormal patterns may merely capture the traits of news forgeries within specific datasets or limited temporal windows,
neglecting the evolution of attributes of fake news.
Besides, attackers could circumvent the pattern analyzers by mimicking the style of real news.

Recent works~\cite{shahi2021overview,vo2018rise, zhang2019multi, ma2019sentence, hardalov2022crowdchecked} \minor{have} prominently recognized \textit{evidence} as a premier element in fake news detection apart from patterns.
This recognition stems from the observation that individuals tend to search for related news as references during the decision-making process.
Researchers have developed many evidence-based FND schemes, where the underlying methodology can be categorized into two types. 
The first is \textit{high-quality dataset contribution}, which provides cleaned news-evidence pairs and the label of the news is determined by whether the majority of conclusions from the evidences match that of the news.
Famous examples of such datasets include PolitiFact ~\cite{rashkin2017truth}, Snopes~\cite{snopes}, etc.
The second is \textit{tailored network design}, which explores enhanced strategies for the news-evidence joint learning~\cite{ma2019sentence, vo2021hierarchical, wu2021unified, wu2021evidence}. 
These methods usually refine and fuse the representations of all evidence articles together with that of the news~\cite{declare,vo2021hierarchical}, somehow resemble learning a boundary from the joint distribution of the news and evidences based on the aforementioned datasets.


Despite the efforts made in evidence-aware FND, these methods either require laborious pre-processing operations to secure relevant and high-quality evidences during both the training and inference stage, or
often utilize all evidences as supplement to the news, 
regardless of their quality and quantity.
In real-world applications, the Internet is capable of providing an exploding amount of similar yet unfiltered materials as evidences for a given news. 
It presents challenges in calculating similarity or manually scrutinizing each piece of evidence for quality assurance.
Furthermore, \minor{is it beneficial to examine as many pieces of evidences as possible? Not always.}
When individuals observe the retrieved \minor{queue of evidences}, a tendency is to \minor{scrutinize it sequentially and decisively: first focusing on the content of the evidences starting from the leading ones}, next \minor{relating each to} the news \minor{with comparing and reasoning}, and finally quitting reading further when they are confident to make a decision.
It motivates us to study ways of more efficient utilization of evidences that devoid of annotating, and ensure resilience to low-quality or less-related retrieved evidences.


We propose a new fake news detection approach with annotation-free evidences and early-termination. 
Our method, \textbf{SEE}, mainly innovates the inference procedure, which includes three main steps.
The first is \textbf{Search}, where we search for online materials using the news as query and directly using their titles as evidences without any annotating or filtering procedure.
The second is \textbf{Examine}, where we encode news and evidences into \minor{representations}, and employ separate transformer\minor{-based} \textit{decoders}~\cite{Transformer} with joint self- and cross-attention for sequential \minor{news-evidence information fusion}. 
The third is \textbf{Early-Termination}, where we equip the model with a shared \textit{confidence assessor} that in each time-step, \minor{i.e., index of the evidence in the queue}, determines either to continue examining more evidences or to provide a prediction with adequate confidence.
Fig.~\ref{fig:intro} depicts the mechanism of our method, in comparison with the previous paradigm of \minor{direct joint news-evidence learning}. 
We design a two-stage training mechanism, where we first train the decoders (or feature extractors) and the ultimate binary classifier for useful representation mining from each time-step.
Then we fix these networks to train the confidence assessors where the target is one minus the distance between the predicted result in each time-step and the label of the news.

To verify the proposed method, we collect evidences via the Microsoft Bing API~\cite{BingAPI} for the news in two famous FND datasets, namely, Weibo21~\cite{MDFEND} and GossipCop~\cite{fakenewsnet}, and refrain from making any additional quality control to the retrieved evidences.
We apply many state-of-the-art methods on the datasets with the same evidences and
experiments show that our method provides the highest average accuracy. 
Besides, we verify the robustness of our SEE approach by alternation, shuffling, removal, or void insertion of evidences, which show little impact on the overall FND performance. 
Moreover, we conduct experiments on two more datasets already with processed evidences, namely, Snopes~\cite{snopes} and PolitiFact~\cite{declare}, which also show leading results. 

The contribution of this paper is three-folded:

\begin{itemize}
\item We propose a fake news detection approach that is capable of \minor{retrieving useful information from} annotation-free online \minor{retrieved} materials, which saves laborious annotating or other pre-processing procedures for quality control over utilized evidences. 
\item We sequentially feed the evidences and devise an early-termination mechanism to use the evidences more efficiently. 
The assessor gives a confidence score in each time-step and determines when to quit reading further evidences and move on to giving a prediction.
\item Experiments on datasets without pre-processed evidences, i.e., Weibo21 and GossipCop, or with pre-processed evidences, i.e., PolitiFact, Snopes, consistently demonstrate the effectiveness of the proposed SEE compared to the state-of-the-art methods.
\end{itemize}

\section{Related Works}
\subsection{Pattern-based Fake News Detection}
Numerous studies have been conducted to develop automatic methods that detect fake news without considering external evidence. 
BiGRU~\cite{ma2016detecting} and TextCNN~\cite{textcnn} respectively use a bidirectional GRU and a 1-D CNN module for feature extraction from the text.
BERT~\cite{BERT} is also frequently utilized as FND baselines where the parameters are kept tunable and the classifier works on the $\operatorname{CLS}$ token.
Ajao et al.~\cite{ajao2019sentiment} propose to analyze the sentimental characteristics of fake news, benefiting some latter works~\cite{zhang2021mining}. 
M$^3$FEND~\cite{M3FEND} uses a memory bank to enrich domain information of the news to assist with detection. 
Also, there are a list of multimodal FND methods that further consider the joint distribution of image and text.
Chen et al.~\cite{WWW} use VAEs to compress the images and texts and learn to minimize the Kullback-Leibler (KL) divergence for correctly matched image-text pairs contrastively.
Ying et al. \cite{yingBoots} extract features from multiple views and design a scoring mechanism to adaptively adjust the weight of each view in the final decision.
Zhou et al.~\cite{zhou2023multi,zhouCLIP} design multi-modal fusion mechanisms with pre-trained models.
Nevertheless, these methods mainly rely on pattern analysis of static news, neglecting the possible evolvement of characteristics of fake news.

\begin{figure*}[!t]
  \centering
  \includegraphics[width=0.99\textwidth]{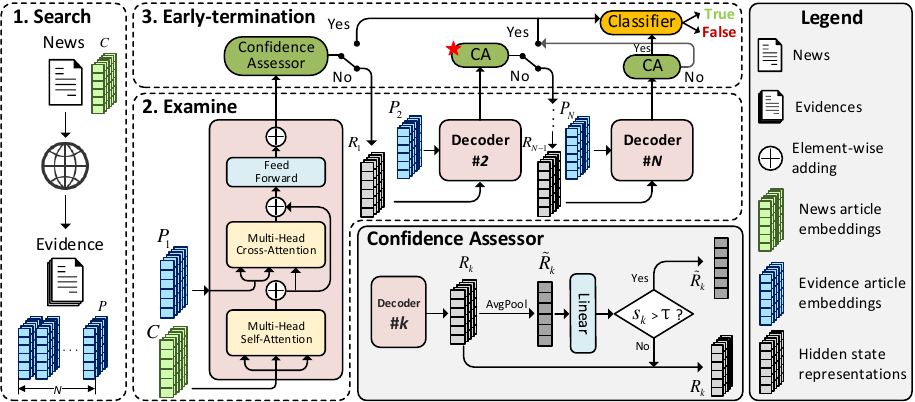}
  \caption{
  Network design of SEE, \minor{our fack-checking FND scheme with annotation-free evidences. SEE includes three main stages:}
\textbf{1. Search}: online materials are retrieved as evidences without any annotating or filtering.
\textbf{2. Examine}: the decoders sequentially examine the news alongside with each piece of evidence via self- and cross-attention to acquire more comprehensive information. 
\textbf{3. Early-termination}: a shared confidence assessor reduces the hidden states in each time-step into confidence scores and determine either to continue examining more evidences or to terminate and predict.}
  \label{fig:architecture}
\end{figure*}

\subsection{Evidence-aware Fake News Detection}
Many high-quality fake news datasets with evidences are proposed.
Snopes~\cite{snopes} and PolitiFact~\cite{rashkin2017truth} contains retrieved evidence articles for each claim by issuing each claim as a query to the Microsoft Bing API, articles are processed by filtering out those related to Snopes and PolitiFact websites and calculating relevance scores to decide on their usage.
Similar datasets are FEVER~\cite{thorne-etal-2018-fever}, Emergent~\cite{ferreira2016emergent}, etc.
In contrast, other famous datasets such as Weibo21 and GossipCop do not provide evidences and therefore have only been applied for pattern-based, rather than evidence-aware, FND.
Besides, many tailored detection networks are proposed on top of these datasets.
DeClarE~\cite{declare} averages signals from external evidence articles and concatenates them with the language of the articles and the trustworthiness of the sources.
Vo et al.~\cite{vo2021hierarchical} proposes to use hierarchical multi-head attention
network to combine word
attention and evidence attention.
CCN~\cite{OpenDomain} leverages both the image caption and text for online evidences, and detects the consistency of the claim-evidence (text-text and image-image), in addition to the image-caption
pairing.
Xu et al.~\cite{xu2022evidence} introduces GET that applies a Graph Neural Network (GNN) to capture long-distance semantic dependency among the news and evidence articles.
However, previous methods either require laborious pre-processing operations towards the evidences, or utilize all evidences for news cases regardless of the quality and quantity.
We propose a new FND approach with annotation-free evidences and early-termination.

\section{Proposed Approach}
Fig.~\ref{fig:architecture} depicts the pipeline of the proposed SEE approach.
It consists of three stages, namely, 1) \textit{S}earching online materials using the news as query and directly using their titles as the retrieved evidences without any annotating or filtering procedure, 2) sequentially \textit{E}xamining the news alongside with each piece of evidence via attention mechanisms, which produce new hidden states with richer information, and 3) using \textit{E}arly-termination within the examination loop by assessing whether sufficient confidence for a correct FND prediction exists.

\subsection{Search: Evidence Preparation}
\label{sec:search}
Let the input news be $\mathcal{N}=[\mathbf{c},\mathcal{E}]\in\mathcal{D}$, where $\mathbf{c},\mathcal{E},\mathcal{D}$ are the text of the news, \minor{a queue} of the corresponding retrieved evidences and the news dataset, respectively.
$\mathcal{E}=[e_1, e_2, ...]$ can be either provided by the dataset along with $\mathbf{c}$, e.g., PolitiFact and Snopes, or prepared by the users via online searching, e.g., Weibo21 and GossipCop.
The searching step can be optional if the evidences can be prepared in advance, otherwise, we prevail the circumvention of laborious annotating or other pre-processing operations on $\mathcal{E}$ for quality control.
\minor{For evidence retrieval, our preliminary is that users trust a certain credibility control of the applied searching engine for collecting evidence through the corresponding API. 
Here, we apply the popular Microsoft Bing to collect evidences. 
To benchmark ``annotation-free'' evidence retrieval, we purposefully exclude any pre-processing operations other than pasting the news content into the search box \& clicking ``Search'' for all datasets.}
We turn each piece of \minor{the retrieved} material into evidence via recording its title, in accordance with Snopes and PolitiFact, and store in order at most $N$ evidences in each $\mathcal{E}$, where $N$ is a tunable hyper-parameter.
Exampled $\mathcal{N}$ for Weibo21 and GossipCop are provided in the experiments\footnote{We will open-source the collected online evidences for Weibo21 and GossipCop after the anonymous reviewing process.}. 
Next, we adopt the BERT model \cite{BERT} as the preliminary feature extractor for both news and the evidences.  
The representations of the news and evidences are respectively denoted as 
$C = \operatorname{BERT}(\mathbf{c}) \in \mathbb{R} ^ {L\times d}, P_i = \operatorname{BERT}(e_i) \in \mathbb{R} ^ {L\times d}$, $i\in [1,N]$, and $d$ denotes the feature dimension.

\subsection{Examine: Attention-based News-evidence Fusion}
We prepare a cascade of $N$ independent transformer decoders~\cite{BERT} with joint self- and cross-attention to sequentially examine the news alongside with each piece of evidence and produce new hidden states with richer information.
Each decoder is responsible for feature fusion for a typical time-step. 
The first block takes $C$ and the leading evidence $P_1$ as inputs, and outputs $R_1$, which we call \textit{hidden state} of the first time-step. 
The following blocks take the hidden state by the previous block and the initial embedding of the upcoming evidence to iteratively produce a list of hidden states. We use the attention mechanism and feed-forward layer operation in typical transformers~\cite{BERT}, denoted with $\operatorname{Attn}(\cdot)$ and $\operatorname{FFN}(\cdot)$. Specifically, the attention operation has:
\begin{equation}
    \operatorname{Attn}(Q,K,V) = \operatorname{softmax}(\frac{QK^T}{\sqrt{d}})V,
\end{equation}
where $Q,K,V$ denotes the query, key, and value matrix.

In the decoders, we combine both self-attention and cross-attention in the examining step for news-evidence interaction.
Each decoder consists of three layers: a self-attention block, a cross-attention block, and a feed-forward network. Each layer has a residual connection to the previous layer and is attached to a layer normalization. For simplicity in equations, we use overlines to denote the Layer Normalization operations (LN)~\cite{layernorm}. The examination process of a decoder block is:
\begin{align}
\begin{split}
L_1 &= 
\begin{cases}
C + \operatorname{Attn}(\overline{C}, \overline{C},\overline{C}), &\emph{if } k=1\\
R_{k-1} + \operatorname{Attn}(\overline{R_{k-1}}, \overline{R_{k-1}},\overline{R_{k-1}}) &\emph{otherwise}
\end{cases}, \\
L_2 &= L_1+ \operatorname{Attn}(\overline{L_1}, \overline{P_k}, \overline{P_k}), \\ 
R_k &= L_2 + \operatorname{FFN}(\overline{L_2}).   
\end{split}
\end{align}
\minor{
Note that each decoder is independent and receives evidence from a specific index in the queue, where the inherent relationship between each index and its statistical relevance of the evidence to the news can be implicitly learned.
One can also alternatively employs a shared decoder and prompts it using different indices to save computation. However, this approach leads to decreased accuracy in comparison to using individual decoders, as shown in the ablation studies (Section~\ref{sec_ablation}).
}

\begin{figure}[!t]
  \centering
  \includegraphics[width=1.0\linewidth]{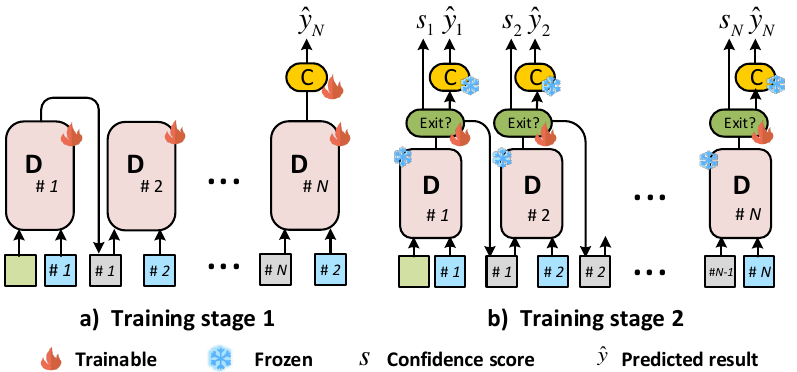}
  \caption{Illustration of the two training stages. In stage 1 we solely emphasize feature extraction from evidences from different index in the \minor{retrieved} queue. In stage 2 the assessor transforms the representations in each time-step into confidence scores for early-exiting on top of fixed feature extractors. The inference stage is depicted in Fig.~\ref{fig:intro}.
  }
  \label{fig:trainingprocedure}
\end{figure}

\subsection{Early-termination: Exit When ``Confident''}
Consider that examining all evidences as supplements to the news regardless of the quality and quantity might be sub-optimal, as the existence of low-quality or less-relevant evidences could impact the model performance.
We introduce the early-termination step that enables the model to cease further examining more pieces of evidence and proceed to prediction if a confidence threshold is surpassed within each time-step. 

In detail, the confidence assessor visits every hidden state $R_k, k\in[1,N]$ and reduces them into confidence scores $s_k$.
We set a hyper-parameter $\tau$ as a threshold. 
$\tau$ is set to decide whether to continue the examining step by sending $R_k$ and the next, if exists, evidence into the next decoder, or to terminate and make the prediction, denoted as $\hat{y}$, by sending $\overline{R_k}$ into the classifier.
$\tau$ is another hyper-parameter, which has a non-negligible effect on both the overall ratio of early-termination and detection performance. 
We give a detailed analysis of $\tau$ in \minor{Section~\ref{section_analysis}}.

During the iterative examining and confidence assessment, we record the hidden state with the highest confidence score, denoted as $R'$.
If the last evidence is reached, we resort to using $R'$ with the highest score to make the prediction $\hat{y}$.
The classifier is a three-layer perceptron and a sigmoid function. Therefore, we denote the calculation procedure of $\hat{y}$ as follows.
\begin{align}
\hat{y} = 
\begin{cases}
\sigma(\operatorname{MLP}(\overline{R_k})), &\emph{if } s_k\ge\tau \\
\sigma(\operatorname{MLP}(\overline{R'})) & \emph{otherwise}
\end{cases}. 
\end{align}

For the confidence assessor, given a hidden state $R_k$ at time-step $k$, we first reduce the token dimension using average pooling over all tokens and use a fully-connected layer $f(\cdot)$ to reduce it into a logit, which is further mapped within $[0, 1]$ as the confidence score by a sigmoid function $\sigma(\cdot)$.
\begin{equation}
\begin{split}
    \overline{R_k} &= \operatorname{AvgPool}(R_k), \\
   s_k &= \sigma(f(\overline{R_k})).
\end{split}
\end{equation}
\noindent

\subsection{Two-staged Training Mechanism}
There is no external label for the confidence assessment process.
\minor{We propose a two-staged training mechanism where we first solely emphasize on feature extraction from evidences, and then train an assessor to determine when to exit with fixed feature extractors. 
Soft labels are automatically assigned in each time-step by comparing the current predicted result with the ground-truth FND hard label.
}

\begin{algorithm}[!t]
\caption{Pseudo-codes for \minor{the} training pipeline.}
\label{algorithm_training}
\raggedright{\textbf{Input}: Training set: $\mathcal{T}$, Validation set: $\mathcal{V}$, Amount of evidences: $N$, Training epochs: $M$}\\
\raggedright{\textbf{Output}: Assessor parameters: $\Omega$, Model parameters besides $\Omega$: $\Theta$}\\
\begin{algorithmic}[1] 
\FOR{m in range($M$):}
\STATE Sample $\{\mathbf{c}, \mathcal{E}, y\}$ from $\mathcal{T}$, $\mathcal{E}=[e_1, ..., e_N]$.
\STATE Extract features $C=\operatorname{BERT}(\mathbf{c})$, $P_k=\operatorname{BERT}(e_k)$.
\STATE // \textit{Stage One: for Feature Extraction \& Detection.}
\STATE Let $R_0=C$.
\FOR{k in range($N$):}
\STATE $R_k=\operatorname{Decoder}_k(R_{k-1}, P_k)$
\ENDFOR
\STATE $\hat{y}=\sigma(\operatorname{MLP}(\operatorname{AvgPool}(R_N)))$ 
\STATE Compute $\mathcal{L}_{cls}(y,\hat{y})$ and update parameters in $\Theta$.
\STATE // \textit{ Stage Two: for Early-Exiting.}
\FOR{k in range($N$):}
\STATE $R_k=\operatorname{Decoder}_k(R_{k-1}, P_k)$
\STATE $s_k = \sigma(f(AvgPool({R_k})))$
\STATE $\hat{y}_n=\sigma(\operatorname{MLP}(\operatorname{AvgPool}(R_k)))$
\STATE $y^{\prime}_k = 1 - \lvert y - \hat{y_k} \rvert$.
\ENDFOR
\STATE Compute $\mathcal{L}_{CA}$ and update parameters in $\Omega$.

\ENDFOR
\end{algorithmic}
\end{algorithm}
\begin{algorithm}[!t]
\caption{Pseudo-codes for \minor{the} inference pipeline.}
\label{algorithm_inference}
\raggedright{\textbf{Input}: Validation/Test set: $\mathcal{V}$, Amount of evidences: $N$, Early-exiting threshold: $\tau$} \\
\raggedright{\textbf{Output}: Prediction: $y$} \\
\begin{algorithmic}[1] 
\STATE Sample $\{t, \mathcal{E}\}$ from $\mathcal{V}$, $\mathbf{E}=[e_1,...,e_N]$.
\STATE Extract features $C=\operatorname{BERT}(t)$, $P_k=\operatorname{BERT}(e_k)$.
\STATE Let $R_0=C$, $s_{max}=0$, $t=0$.
\FOR{k in range($N$):}
\STATE $R_k=\operatorname{Decoder}_k(R_{k-1}, P_k)$
\STATE $s_k = \sigma(f(\operatorname{AvgPool}(R_k)))$
\IF{$s_k > \tau$ }
\RETURN $y=\sigma(\operatorname{MLP}(\operatorname{AvgPool}(R_k)))$
\ELSIF{$s_k>s_{max}$}
\STATE Let $s_{max}=s_k$, $t=k$.
\ENDIF
\ENDFOR
\RETURN $y=\sigma(\operatorname{MLP}(\operatorname{AvgPool}(R_t)))$
\end{algorithmic}
\end{algorithm}

\noindent\textbf{Stage One: Training Feature Extractors and Classifier.} 
The initial phase involves training the feature extractors, i.e. the tunable BERT and decoders, and the classifier. The confidence assessor remains uninvolved during this stage.
Specifically, we iteratively feed all $N$ evidences without early-termination, and every decoder generates a unique hidden state in each time-step.
The goal is to enable the decoders to extract useful features from the provided news and evidences at different time-steps.
The hidden state of the last decoder $R_N$ is then fed to the classifier, which produces $\hat{y}_N$. We update the networks using the classification loss $\mathcal{L}_{cls}$, where
\begin{equation}
\begin{split}
  \mathcal{L}_{cls} &= y\log( \hat{y}_N)+(1-y)+(1-y)\log(1- \hat{{y}}_N). 
\end{split}
\end{equation}
The classifier is also jointly trained in the settings where all $N$ evidences are provided.
The purpose of this stage is to provide the classifier with adequate information for decision-making as well as encourage the decoders to maximally reserve the most critical information till the last time-step. 

\noindent\textbf{Stage Two: Training the Confidence Assessor.} 
After the first stage, we train the confidence assessor to provide confidence scores for each time-step. 
We fix the models updated in the first stage, i.e., the tunable BERT, the decoders, and the classifier, then only train the confidence assessors where the target is one minus the distance between the predicted result in each time-step and the label of the news.
Note that here a shared assessor is employed across all time-steps in that we want the assessor to terminate inference \minor{on feeling ``confident'' without the prior of index. Here, the ``confidence'' is represented by a predicted score that might not align with human-defined confidence.}
Specifically, we get the predicted scores $\{\hat{y}_1,...,\hat{y}_N\}$ in each time-step by respectively sending the corresponding hidden states $\{\mathbf{R}_1,...,\mathbf{R}_N\}$ into the fixed classifier. 
The labels of the confidence scores, denoted as $y^{\prime}_k$, are defined as one minus the distance between $\hat{y}_k$ and $y$, as in Eq.~\ref{eq:cal_score_gt}, i.e., \minor{the larger the prediction deviates from the ground truth, the lower the confidence should be, ranging from zero to one.}
\begin{equation}
  y^{\prime}_k = 1 - \lvert y - \hat{y}_k \rvert, k\in[1,N].
  \label{eq:cal_score_gt}
\end{equation}
\noindent
The confidence assessor is trained to regress $y^{\prime}_k$ in each time-step, where we use the summed $L_1$ loss for training:

\begin{equation}
  \mathcal{L}_{CA} = \sum^N_{k=1}{\lvert y^\prime_k - s_k\rvert}.
\end{equation}

Finally, we provide the pseudo-code of detailed training and inference processes of SEE in Algo.~\ref{algorithm_training} and Algo.~\ref{algorithm_inference}.

\section{Experiments}
\subsection{Experimental Setups}
\noindent\textbf{Implementation Details.}
We use \emph{bert-base-chinese} and \emph{bert-base-uncased} pre-trained models for processing Chinese dataset and English datasets, respectively. 
The hidden size of word embeddings is 768 ($d=768$). 
We unify the length of input news and evidence to a specific length by padding or truncating ($L=100$). For samples with fewer than $N$ evidence, the missing evidences are treated as blank and filled with [PAD] tokens.
To save computation, the representations can be pre-computed, stored on disk and loaded upon training.
Our model is trained using a single NVIDIA GeForce RTX 4090 GPU. We use Adam optimizer \cite{adam} with default parameters. The batch size is 12. 
We use a learning rate of $6 \times10^{-6}$ for the feature extractors, i.e., fine-tuned BERT and the decoders, and $5\times10^{-5}$ for the rest of the model. 

\noindent\textbf{Data Preparation.} 
We use four famous datasets, namely Weibo21~\cite{MDFEND}, GossipCop~\cite{fakenewsnet}, Snopes~\cite{snopes}, and PolitiFact~\cite{declare} for our experiments. 
Weibo21 is a Chinese fake news detection dataset collected from Sina Weibo.
\minor{News content of} GossipCop,
Snopes, 
and PolitiFact 
are collected from fact-checking websites. 
Weibo21 and GossipCop do not contain official evidences, so we collect evidence articles for them by the method mentioned in Section~\ref{sec:search}.
We conduct train-val-test split in the ratio of 6:2:2 in accordance with previous methods. 
Table~\ref{tab:datasets} summarizes these datasets.

\begin{table}[!t]
\caption{Statistics of the applied FND datasets. \# denotes ``the number of".}
\centering
  \begin{tabular}{lrrrr}
  \toprule
    \textbf{Dataset} & \textbf{\# Total} & \textbf{\# True} & \textbf{\# False} & \textbf{\# Evidence} \\ \midrule
    Weibo21 & 9,128 & 4,640 & 4,488 & 90,550 \\ 
    GossipCop & 22,140 & 16,817 & 5,323 & 218,525 \\
    Snopes & 4,341 & 1,164 & 3,177 & 29,242 \\
    PolitiFact & 3,568 & 1,867 & 1,701 & 29,556 \\
    \bottomrule
  \end{tabular}
  \label{tab:datasets}
\end{table}


\begin{figure}[!t]
  \centering
  \includegraphics[width=0.98\linewidth]{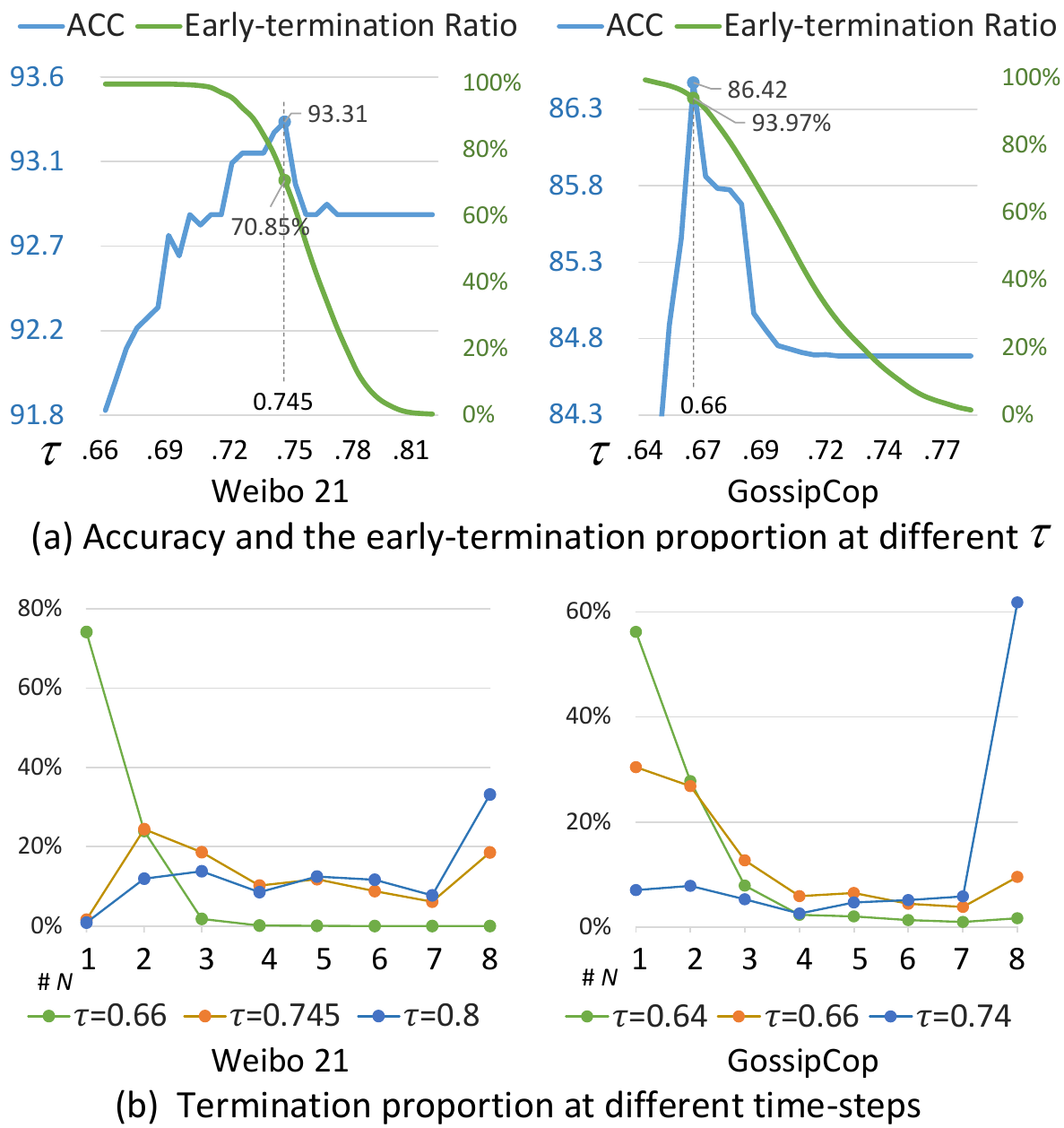}
  \caption{Quantitative analysis of the impact of $\tau$ on the performance and early-termination. 
  }
  \label{fig:experiment1}
\end{figure}

\subsection{Performance Analysis}
\label{section_analysis}
\minor{Before conducting extensive comparisons with previous state-of-the-arts, we first study the impact of different implementation settings and evidence arrangements.}
\minor{By doing this we aim to elaborate on the selections of key parameters and their impact on the performances, as well as verifying the stability of SEE under different evidence-retrieving situations.}

\noindent\textbf{Impact of the Threshold $\tau$ on Accuracy \minor{and Efficiency}.} 
The threshold $\tau$ influences early-termination and therefore also accuracy \& efficiency of our method.
In the row (a) of Fig.~\ref{fig:experiment1}, we record the detection accuracy alongside with the early-termination rate using varied values of $\tau$. 
\minor{Here the early-termination ratio calculates the probability of over-threshold confidence scores, which is defined as the number of samples that terminate early divided by the total number of samples.}
For accuracy curves (blue lines), they first rise with the increase of $\tau$, and then descend and stabilize after reaching the peak value.
This observation shows that using information from adequate pieces of evidence can be often enough to produce a reliable prediction, which \minor{might be counter-intuitive with the point that viewing all of the retrieved evidences is essential and better.}


\begin{table*}[!t]
\caption{Quantitative analysis of the impact of \minor{the retrieved} results from the view of quality, quantity, and order, detailed description of each setting is provided in section~\ref{section_analysis}. We report the accuracy, macro F1 score, and area under ROC (AUC). *: using the pre-filtered evidences within the dataset.}
\centering
\label{tab:quantitative}
\renewcommand{\arraystretch}{1.2}
\setlength{\tabcolsep}{0.9mm}{
  \begin{tabular}{m{3.8cm}|ccc|ccc|ccc|ccc}
  \toprule
  \multirow{2}{*}{\centering\textbf{Evidence Adjustments}} & \multicolumn{3}{c|}{Weibo21} & \multicolumn{3}{c|}{GossipCop} & \multicolumn{3}{c|}{Snopes*} & \multicolumn{3}{c}{PolitiFact*}
  \\
  \cline{2-13}
 & ACC & F1 & AUC & ACC & F1 & AUC & ACC & F1 & AUC & ACC & F1 & AUC \\
     \midrule
   Most Related Swapped & 0.927 & 0.927 & 0.927 & 0.860 & 0.798 & 0.783 & 0.819 & 0.774 & 0.766 & 0.663 & 0.659 & 0.662 \\ 
   All Evidences Shuffled & 0.926 & 0.926 & 0.926 & 0.856 & 0.792 & 0.776 & 0.798 & 0.747 & 0.740  & 0.667 & 0.664 & 0.666 \\ 
   Reverse the Sequence & 0.927 & 0.927 & 0.927 & 0.860 & 0.768 & 0.782 & 0.798 & 0.747 & 0.739  & 0.676 & 0.671 & 0.674 \\ 
   Most Related Void & 0.931 & 0.931 & 0.930 & 0.860 & 0.796 & 0.780 & 0.806 & 0.748 & 0.732 & 0.646 & 0.632 & 0.644 \\ 
   Most Related Missing & 0.926 & 0.926 & 0.925 & 0.857 & 0.787 & 0.765 & 0.746 & 0.667 & 0.657 & 0.608 & 0.603 & 0.607 \\
   Limited Evidence (1) & 0.905 & 0.905 & 0.904  & 0.852 & 0.783 & 0.765 & 0.770 & 0.721 & 0.719  & 0.676 & 0.671 & 0.674 \\ 
   Limited Evidences (3) & 0.926 & 0.926 & 0.926  & 0.857 & 0.788 & 0.767 & 0.812 & 0.758 & 0.743  & 0.694 & 0.691 & 0.693 \\ 
   \textbf{Baseline Setting} & \textbf{0.932} & \textbf{0.932} & \textbf{0.932} & \textbf{0.864} & \textbf{0.807} & \textbf{0.796} & \textbf{0.824} & \textbf{0.786} & \textbf{0.783} & \textbf{0.706} & \textbf{0.705} & \textbf{0.706} \\ 
    \bottomrule
  \end{tabular}
  }
\end{table*}

Meanwhile, $\tau$ also influences the efficiency of inference, where larger $\tau$ indicates that the model would be more cautious during examinations, empirically requiring more evidences on average to produce a determination.
We visualize the proportion of termination at each time-step in the row (b) of Fig.~\ref{fig:experiment1}. 
The orange curve illustrates the optimal $\tau$ on the dataset \minor{if we only prevail high accuracy.}
It generally manifests an even distribution, suggesting that the leading evidences only gain moderate extra importance in comparison with the rest of evidences. 

\begin{table*}[!t]
\caption{Performance comparison between our method with others. *: using the pre-filtered evidences within the dataset. $^1$:~methods not using evidences, $^2$:~methods using evidences. $-$:~lacking domain information for experiments.}
\centering
\label{tab:comparison}
\centering
\begin{tabular}{l|ccc|ccc|ccc|ccc}
\toprule
\multirow{2}{*}{\textbf{Method}} &\multicolumn{3}{c|}{Weibo21} & \multicolumn{3}{c|}{GossipCop} & \multicolumn{3}{c|}{Snopes*} & \multicolumn{3}{c}{PolitiFact*}\\   \cline{2-13}
 & ACC & F1 & AUC & ACC & F1 & AUC & ACC & F1 & AUC & ACC & F1 & AUC\\
\midrule
BiGRU$^1$~\cite{ma2016detecting} & 0.827 & 0.827 & 0.898 & 0.781 & 0.783 & 0.764 & 0.712 & 0.615 & 0.608 & 0.624 & 0.623 & 0.623\\
TextCNN$^1$~\cite{textcnn} & 0.872 & 0.872 & 0.873 & 0.840 & 0.757 & 0.734 & 0.697 & 0.502 & 0.529 & 0.592 & 0.592 & 0.592\\
BERT$^1$~\cite{BERT} & 0.918 & 0.918 & 0.918 & 0.855 & 0.792 & 0.779 & 0.733 & 0.627 & 0.619 & 0.604 & 0.583 & 0.601\\
M$^3$FEND$^1$~\cite{M3FEND} & 0.922 & 0.922 & \textbf{0.975} & 0.824 & - & - & - & - & - & - & - & -\\
DeClarE$^2$~\cite{declare}  & 0.850 & 0.850 & 0.850 & 0.798 & 0.670 & 0.667 & 0.786 & 0.694 & 0.677 & 0.635 & 0.630 & 0.631 \\
CCN-sent$^2$~\cite{OpenDomain} & 0.853 & 0.852 & 0.853 & 0.854 & 0.783 & 0.764 & 0.740 & 0.658 & 0.649 & 0.677 & 0.676 & 0.676 \\
CCN-lstm$^2$~\cite{OpenDomain} & 0.876 & 0.875 & 0.875 & 0.826 & 0.747 & 0.734 & 0.747 & 0.684 & 0.679 & 0.624 & 0.623 & 0.625\\
GET$^2$~\cite{xu2022evidence} & 0.666 & 0.662 & 0.667 & 0.847 & 0.773 & 0.754 & 0.814 & 0.771 & 0.767 & 0.694 & 0.689 & 0.692 \\
\textbf{SEE$^2$ (Proposed)}  & \textbf{0.932} & \textbf{0.932} & 0.932 & \textbf{0.864} & \textbf{0.807} & \textbf{0.796} & \textbf{0.824} & \textbf{0.786} & \textbf{0.783} & \textbf{0.706} & \textbf{0.705} & \textbf{0.706}  \\
  \bottomrule
 \end{tabular}
\end{table*}

\noindent\textbf{Selection of proper $\tau$.} 
We search for the best early-termination threshold $\tau$ on the validation set before testing. 
\minor{In the real world, the performance of overall detection accuracy is usually in favored of over inference time, so here we temporarily select $\tau$ without additionally considering the trade-off between accuracy and efficiency.}
According to Fig.~\ref{fig:experiment1}, $\tau=0.745$ and $\tau=0.660$ respectively yield the best accuracy on the validation set of Weibo21 and GossipCop. 
Besides, the optimal threshold for Snopes is $\tau=0.715$, and that for PolitiFact $\tau=0.690$.
In practical scenarios, since the four applied datasets are representative and popular, users can select a proper $\tau$ ranging from $0.715$ to $0.745$ for Chinese news, or $0.660$ to $0.690$ for English news. 
\minor{Moreover, the experiments indicate that even when the threshold is roughly set within an acceptable small range, the caused variation in accuracy is within $2\%$ of the peak value for both datasets.}

\noindent\textbf{Impact of \minor{the Retrieved} Results.}
\minor{Provided with the same query, the searching engine might also retrieval different combinations of materials over time, potentially changing the prediction results of fact-checking based FND methods. Therefore, after the selection of proper $\tau$,} we also study the impact of \minor{the retrieved} results from the view of quality, quantity, and order. 
The settings are:
1) \textit{Most Related Swapped} presents the leading three \minor{(usually most related) materials in the retrieved queue in an arbitrary order, while keeping intact} the remaining materials.
2) \textit{All Evidences Shuffled} simulates a more challenging scenario where all materials are in \minor{an arbitrary} order.
3) \textit{Reverse the Sequence} simulates a scenario in which less relevant evidence is examined first, which is an even more challenging scenario compared to \textit{All Evidences Shuffled}.
3) \textit{Most Related Void} simulates that the first (primary) evidence conveys no valuable information, e.g., due to advertisement or \minor{failed retrieval}. In this case the first evidence is replaced with a random irrelevant article.
4) \textit{Most Related Missing} simulates the absence of the primary evidence. \minor{In this case the first evidence is simply polled out from the queue.}
5) \textit{Limited Evidences ($n$)} simulates that the \minor{searching queue only contains} a maximum of $n$ evidence.
\minor{On doing the experiments, we first train our model using the default content and order of evidences in each dataset, and then test them using the above-mentioned settings.}

The results are reported in Table~\ref{tab:quantitative}.
\minor{Among all adjustments, providing \textit{limited evidences (1 or 3)} decreases performance most evidently, which suggests that considering more evidence is necessary for detection.} 
Adjustments on the order of evidences or the content of the most related evidences result in different \minor{performance drop trend} on different datasets. 
For datasets without pre-processed evidences, i.e., Weibo21 and GossipCop, \minor{we see that the performance drop is trivial, i.e., usually less than 1\%. 
In contrast, the drop is more noticeable on datasets with pre-processed evidences, i.e., Snopes and PolitiFact, even if only the leading materials are swapped.
The finding suggest that the pre-filtering stage might has the ability to indeliberately mislead the model to over-highlight the order of the evidences as well as the role played by the most related evidence.}
\minor{In comparison, using annotation-free evidences suggest a more robust and consistent result less affected by the quality of the first evidence and the order.}


\subsection{Comparisons}
The compared benchmark methods are listed in Table~\ref{tab:comparison}. 
\minor{
Since our testing datasets might not have been used by some of the baseline methods, we carefully re-implemented them by providing all available required data, e.g., publisher information, propagation graph, etc. In contrast, SEE refrains from using the related additional information other than annotation-free evidences.
}

\noindent\textbf{Implementation of Baselines.}
\minor{
DeClare~\cite{declare} and GET~\cite{xu2022evidence} also require publisher information \minor{and other propagation-based information.
As such,} we do not directly borrow the experimental results from their papers and instead collect related information and carefully retrain the models on the testing datasets.
GET involves generating graphs by segmenting words and using pre-trained embeddings, and we implement this on Weibo21 by using jieba~\footnote{https://github.com/fxsjy/jieba} and Chinese word vectors pre-trained on weibo~\cite{weiboembedding}.
}
Domain information of news is mandatory for M$^3$FEND, which is not available in GossipCop, Snopes, and PolitiFact.
We remove visual modality parts of the CCN\minor{~\cite{OpenDomain}} and report the results using Sentence-BERT~\cite{SentenceBert} and pre-trained BERT with LSTM versions of it, denoted respectively as CCN-sent and CCN-lstm. 

\noindent\textbf{Results.}
Table~\ref{tab:comparison} presents accuracy, macro F1 score, and area under ROC (AUC) for performance evaluation. 
SEE achieves the average accuracy on all four datasets, which outperforms the compared benchmark methods. 
SEE also attains the highest F1 score across all datasets, demonstrating its classification ability.
GET exhibits competitive performance with SEE on Snopes and PolitiFact. 
However, SEE outperforms it notably on datasets without pre-processed evidences, i.e., Weibo21 and GossipCop. 
Similarly, CCN performs closely to SEE on datasets without pre-processed evidences, yet underperforms on Snopes and PolitiFact.
The reason \minor{is mainly that}
GET and CCN are proposed to examine evidence alongside with news at levels of word without information retrieval stage, which demands a high quality of evidence text. 
\minor{As a result, the models are not trained explicitly to} utilize un-filtered evidences, \minor{which might show different characteristics with the filtered ones}.
\minor{Notably, we see that these evidence-based methods might not significantly outperform methods without evidence on Weibo21 and GossipCop, but both show decent performance gain on datasets with pre-processed evidence.
Therefore, we could infer that the performance boost of these methods relies partially on the pre-processing stage in these methods.}
In contrast, the consistently strong performances of SEE on four datasets demonstrate that it overcomes the above-mentioned disadvantages, which could circumvent the time-consuming human-aided filtering.

\subsection{Ablation Studies}
\label{sec_ablation}
\noindent\textbf{Enabling Different Amount of Evidences.} In Table~\ref{tab:ablation}, we vary the maximum quantities of evidences for testing. 
The performances drop evidently \minor{compared to the default setting}.
It suggests that the training stage is impacted altogether by the \minor{quantity of evidences, even though some of which might show less relation with the news.} 
\minor{The same performance drop happens to results on Snopes and PolitiFact, suggesting this does not depend on evidence quality}. 
As stage one considers evidences sequentially, and produces detection outcomes solely at the final decoder, an excessive input-to-output span thus damages the training. The input of front decoders may be overshadowed as the data propagates deeper. 
Therefore, limiting the number of input evidence, which equals limiting the depth of SEE, benefits the performance.


\begin{table}[!t]
\caption{Ablation studies of the proposed SEE on network design and training mechanism.}
\label{tab:ablation}
\centering
\renewcommand{\arraystretch}{1.2}
\setlength{\tabcolsep}{0.6mm}{
  \begin{tabular}{l|ccc|ccc}
  \toprule
  \multirow{2}{*}{\centering\textbf{Ablation Settings}} & \multicolumn{3}{c|}{Weibo21} & \multicolumn{3}{c}{GossipCop}
  \\
  \cline{2-7}
 & ACC & F1 & AUC & ACC & F1 & AUC \\
     \midrule
   Max. allow 6 evidences & 0.922 & 0.922 & 0.922 & 0.858 & 0.788 & 0.766   \\ 
   Max. allow 10 evidences & 0.916 & 0.916 & 0.916 & 0.854 & 0.779 & 0.755  \\ 
   Decoders sharing weights & 0.883 & 0.883 & 0.884 & 0.847 & 0.771 & 0.749 \\ 
   Training models in one go & 0.827 & 0.827 & 0.898 & 0.781 & 0.783 & 0.764 \\
   Concat All Hidden States & 0.929 & 0.928 & 0.928  & 0.859 & 0.798 & 0.781 \\ 
   \textbf{Baseline setting} (Max. 8) & \textbf{0.932} & \textbf{0.932} & \textbf{0.932} & \textbf{0.864} & \textbf{0.807} & \textbf{0.796} \\ \midrule
  \multirow{2}{*}{\centering\textbf{Ablation Settings}} & \multicolumn{3}{c|}{Snopes} & \multicolumn{3}{c}{PolitiFact}
  \\
  \cline{2-7}
 & ACC & F1 & AUC & ACC & F1 & AUC \\
     \midrule
   Max. allow 6 evidences & 0.821 & 0.778 & 0.772 & 0.704 & 0.484 & 0.524   \\ 
   Max. allow 8 evidences & 0.751 & 0.592 & 0.595 & 0.636 & 0.597 & 0.611  \\ 
   Decoders sharing weights & 0.780 & 0.720 & 0.710 & 0.664 & 0.662 & 0.665 \\ 
   Training models in one go & 0.803 & 0.748 & 0.736 & 0.686 & 0.681 & 0.682  \\
   Concat All Hidden States & 0.818 & 0.785 & 0.763  & 0.701 & 0.696 & 0.695 \\ 
   \textbf{Baseline Setting} (Max. 5) & \textbf{0.824} & \textbf{0.786} & \textbf{0.783} & \textbf{0.706} & \textbf{0.705} & \textbf{0.706} \\ 
    \bottomrule
  \end{tabular}
  }
\end{table}

\noindent\textbf{Shared or Independent Decoders.} In Table~\ref{tab:ablation}, we investigate the utilization of a shared decoder informed by time-step information through positional encodings.
The resulting accuracy witnesses declines of $4.9\%$ and $1.7\%$, $4.4\%$, and $4.2\%$, on four datasets respectively, suggesting that assigning a single decoder to capture universal features across evidences sequential locations compromises the examination proficiency of the proposed SEE.

\noindent\textbf{Two-staged or One-go Training.} We test the necessity of the two-staged training by altering it with \textit{training the model in one go}, in which we jointly train all components.
During training, the hidden state at each time-step is directed into the assessor. In instances where termination is deemed appropriate, \minor{the remaining evidences are disregarded, leaving subsequent decoders untrained. }
The proposed approach suffers from performance drops under this alternative approach on all datasets. 
According to our observation of details, a substantial number of samples tend to terminate right after the first decoder block, indicating that the confidence assessment holds limited validity. We conclude that jointly training all components damages both the examination and termination abilities of SEE.

\noindent\textbf{Concatenating All Hidden States \minor{from Every Time-step.}} 
Previous methods mainly utilize concatenation to fuse representations of examined evidences. We exclude the confidence assessment during inference by concating the hidden states at all time-steps, which produces a hidden representation of size $L\times Nd$. This setting underperforms baseline settings, which demonstrates the effectiveness of sequential examination of evidence.

\section{Conclusions and Future Works}
We introduce SEE, a FND method with Search, Examine and Early-termination \minor{based on annotation-free evidences}.
Our approach incorporates confidence assessment trained on annotation-free evidences for early-termination within examination loops \minor{without the effort of human-intervened evidence labelling.
Through the method, we are motivated by two key insights, which are then verified by experiments: 1) it is not always necessary to utilize as much evidences as possible to make correct FND prediction, and 2) training models upon well-crafted useful information might mistakenly delude it to highlight the order of all evidences as well as the role played by the most relevant one. 
}
Our extensive results underscore the superiority of SEE over state-of-the-art methods, validating its robustness across diverse scenarios. We conclude that SEE excels in distinct evidences utilization and detection capabilities, suggesting that guiding models to assess annotation-free evidences aids in evidence-aware FND.

\minor{
While we made progress in directly utilizing web-searched raw evidences, the in-depth mechanism of evidence examination as well as that of the early-exiting remains implicit to us. It would be beneficial to further improve the interpretability of FND methods by perhaps utilizing large language models which have zero-shot reasoning abilities for a Q\&A.}



\begin{ack}
This work was supported by the National Natural Science Foundation of China under Grants U20B2051, 62072114, U20A20178, and U22B2047.
\end{ack}



\bibliography{mybibfile}

\end{document}



\begin{frontmatter}


\paperid{130} 


\title{Supplemental Materials for "Search, Examine and Early-Termination: Fake News Detection with Annotation-Free Evidences"}


\author[A]{\fnms{First}~\snm{Author}\orcid{....-....-....-....}\thanks{Corresponding Author. Email: somename@university.edu.}\footnote{Equal contribution.}}
\author[B]{\fnms{Second}~\snm{Author}\orcid{....-....-....-....}\footnotemark}
\author[B,C]{\fnms{Third}~\snm{Author}\orcid{....-....-....-....}} 

\address[A]{Short Affiliation of First Author}
\address[B]{Short Affiliation of Second Author and Third Author}
\address[C]{Short Alternate Affiliation of Third Author}


\end{frontmatter}

\section{Experimental Settings}
\subsection{Datasets Settings}
The original PolitiFact dataset assigns each claim to one of the six possible labels: \emph{ture, mostly true, half true, mostly false, false, pants-on-fire}. Following previous works~\cite{declare, xu2022evidence}, we merge \emph{true, mostly true, half true} as the positive class, and \emph{mostly false, false, pants-on-fire} as the negative class for PolitiFact. The publisher information of news is discarded for all datasets.

We divide all the datasets into training, validation, and testing sets in the same ratio of 6:2:2 while maintaining the proportion of positive and negative samples. All benchmark methods as well as SEE are evaluated on the same divided datasts. The specific statics of the divided datasets are presented in Table~\ref{tab:divdatasets}.

\begin{table}[!h]
\small
\centering
  \begin{tabular}{cccccc}
  \toprule
    \multicolumn{2}{c}{\textbf{Dataset}} & \textbf{WB} & \textbf{GC} & \textbf{SN} & \textbf{PF} \\ \midrule
    \multirow{2}{*}{Train} & News & 5480 & 13288 & 2604 & 2140 \\ & Evidences & 54372 & 131122 & 17650 & 16551 \\ \midrule
    \multirow{2}{*}{Val.} & News & 1824 & 4426 & 869 & 713 \\ & Evidences & 18114 & 43751 & 5857 & 6448 \\ \midrule
    \multirow{2}{*}{Test} & News & 1824 & 4426 & 868 & 715 \\ & Evidences & 18107 & 43684 & 5735 & 6557 \\
    \bottomrule
  \end{tabular}
  \caption{Statistics of the divided FND datasets. (WB: Weibo21, GC: GossipCop, SN: Snopes, PF: PolitiFact)}
  \label{tab:divdatasets}
\end{table}

\begin{algorithm}[!ht]
\caption{Pseudo-codes for single-stage training.}
\label{algo:singel_stage}
\textbf{Input}: Training set $\mathcal{T}$, Amount of evidences: $N$, Training epochs: $M$, Early-termination threshold $\tau$\\
\textbf{Output}: Model parameters: $\Theta$.
\begin{algorithmic}[1] 
\FOR{m in range($M$):}
\STATE Sample $\{\mathbf{c}, \mathcal{E}, y\}$ from $\mathcal{T}$, $\mathcal{E}=[e_1, ..., e_N]$.
\STATE Extract features $C=\operatorname{BERT}(\mathbf{c})$, $P_k=\operatorname{BERT}(e_k)$.
\STATE Let $R_0=C$.
\FOR{k in range($N$):}
\STATE $R_k=\operatorname{Decoder}_k(R_{k-1}, P_k)$
\STATE $s_k = \sigma(f(AvgPool({R_k})))$
\STATE $\hat{y}_k=\sigma(\operatorname{MLP}(\operatorname{AvgPool}(R_k)))$
\STATE $y^{\prime}_k = 1 - \lvert y - \hat{y_k} \rvert$.
\IF{$s_k>\tau$}
\STATE $\hat{y}=\hat{y}_k$ 
\STATE Break.
\ENDIF
\ENDFOR
\STATE Compute $\mathcal{L}_{total}$ and update parameters in $\Theta$.
\ENDFOR
\end{algorithmic}
\end{algorithm}

\subsection{Experimental Details}
\begin{itemize}
\item \textbf{BiGRU \cite{ma2016detecting}} consists of a one-layer bidirectional GRU layer with a hidden size of 300, and a three-layer MLP is utilized to classify fake news.
\item \textbf{TextCNN \cite{textcnn}} is a convolution module for text encoding. 
We implement a TextCNN module followed by a three-layer MLP classifier. The convolution module has 5 kernels with the same 64 channels that have steps of 1, 2, 3, 5, and 10 respectively.
\item \textbf{BERT \cite{BERT}} fine-tunes the pre-trained BERT models, and a three-layer MLP classifier utilizes the $[\text{CLS}]$ token for detection.
\item\textbf{M3FEND \cite{M3FEND}} utilizes a memory bank to enrich domain information of the news for detection. We borrow the performance metrics from the original paper.
\item\textbf{DeClarE \cite{declare}} employs BiLSTM and attention mechanism to generate the final representation using news and each word in evidences. We implement the model following the settings reported in their original paper using PyTorch since there is no public code. 
\item\textbf{CCN \cite{OpenDomain}}. The model consists of a pre-trained BERT model or Sentence-bert \cite{SentenceBert} for encoding and a memory network \cite{MemoryNetwrok} to evaluate the consistency of the news against the evidence.
\item\textbf{GET \cite{xu2022evidence}} is a graph-based framework for semantic structure mining. We retrain the model by sticking strictly to the experimental settings of the original paper.
\end{itemize}

We use the embeddings of pre-trained BERT for BiGRU, TextCNN in accordance with SEE. 
We didn't borrow the experimental results from the original paper of DeClare or GET as we discard the publisher information and to ensure a fair comparison on the same dataset partitioning.
Since GET involves generating graphs by segmenting words and using pre-trained embeddings, we implement this on Weibo21 by using jieba~\footnote{https://github.com/fxsjy/jieba} and Chinese word vectors pre-trained on weibo~\cite{weiboembedding}.
We apply early-stoping strategy for experiments by choosing the model parameters that tested optimal on the validation set for final testing.

\subsection{Single-stage Training Strategy}
We implement a training strategy for SEE using only a single stage, in which we train all the components of SEE jointly. 
For each sample during training, we sequentially generate $R_k, s_k$ at each time-step. However, this procedure breaks once the $s_k$ is greater than $\tau$. 
The total loss is calculated using the ultimate prediction $\hat{y}$ and all confidence scores $si$ before breaking. 
Suppose the prediction is produced at the time-step $n$ using hidden state $R_n$, the total loss is formulated as such:
\begin{equation}
    \mathcal{L}_{total}=\mathcal{L}_{cls}(y,\hat{y})+\sum^n_{i=1}\lvert y'_i-s_i \rvert.
\end{equation}  
The pseudo-code of the detailed process of single-stage training is presented in Algo.~\ref{algo:singel_stage}

\section{Experimental Data}

\begin{table}[!t]
\caption{Ablation studies of the proposed SEE on network design and training mechanism.}
\label{tab:ablation}
\centering
\renewcommand{\arraystretch}{1.2}
\setlength{\tabcolsep}{0.6mm}{
  \begin{tabular}{l|rrr|rrr}
  \toprule
  \multirow{2}{*}{\centering\textbf{Ablation Settings}} & \multicolumn{3}{c|}{Weibo21} & \multicolumn{3}{c}{GossipCop}
  \\
  \cline{2-7}
 & ACC & F1 & AUC & ACC & F1 & AUC \\
     \midrule
   Max. allow 6 evidences & 0.922 & 0.922 & 0.922 & 0.858 & 0.788 & 0.766   \\ 
   Max. allow 7 evidences & 0.927 & 0.927 & 0.927  & 0.853 & 0.790 & 0.778  \\ 
   Max. allow 9 evidences & 0.918 & 0.918 & 0.918 & 0.860 & 0.797 & 0.790  \\
   Max. allow 10 evidences & 0.916 & 0.916 & 0.916 & 0.854 & 0.779 & 0.755  \\ 
   Decoders sharing weights & 0.883 & 0.883 & 0.884 & 0.847 & 0.771 & 0.749 \\ 
   Training models in one go & 0.827 & 0.827 & 0.898 & 0.781 & 0.783 & 0.764 \\
   Concat All Hidden States & 0.929 & 0.928 & 0.928  & 0.859 & 0.798 & 0.781 \\ 
   \textbf{Baseline setting} (Max. 8) & \textbf{0.932} & \textbf{0.932} & \textbf{0.932} & \textbf{0.864} & \textbf{0.807} & \textbf{0.796} \\ \midrule
  \multirow{2}{*}{\centering\textbf{Ablation Settings}} & \multicolumn{3}{c|}{Snopes} & \multicolumn{3}{c}{PolitiFact}
  \\
  \cline{2-7}
 & ACC & F1 & AUC & ACC & F1 & AUC \\
     \midrule
   Max. allow 6 evidences & 0.821 & 0.778 & 0.772 & 0.704 & 0.484 & 0.524   \\ 
   Max. allow 7 evidences & 0.795 & 0.710 & 0.689 & 0.694 & 0.573 & 0.571  \\ 
   Max. allow 8 evidences & 0.751 & 0.592 & 0.595 & 0.636 & 0.597 & 0.611  \\ 
   Decoders sharing weights & 0.780 & 0.720 & 0.710 & 0.664 & 0.662 & 0.665 \\ 
   Training models in one go & 0.803 & 0.748 & 0.736 & 0.686 & 0.681 & 0.682  \\
   Concat All Hidden States & 0.818 & 0.785 & 0.763  & 0.701 & 0.696 & 0.695 \\ 
   \textbf{Baseline Setting} (Max. 5) & \textbf{0.824} & \textbf{0.786} & \textbf{0.783} & \textbf{0.706} & \textbf{0.705} & \textbf{0.706} \\ 
    \bottomrule
  \end{tabular}
  }
\end{table}

\subsection{Ablation Studies}
The complete ablation studies results on four datasets are exhibited in Table~\ref{tab:ablation}. The settings of \textit{Max. allow $N$ evidences} are different from the settings in the main paper as these two datasets have different statistical information on evidences. In Snopes and PolitiFact, each news is with seven and eight pieces of evidence on average respectively. However, the medians of these counts are five and five. Therefore, we only choose the maximum usage of six, seven, and eight evidences for experiments.

\subsection{Case Studies}
\begin{figure*}[!t]
  \centering
  \includegraphics[width=1\linewidth]{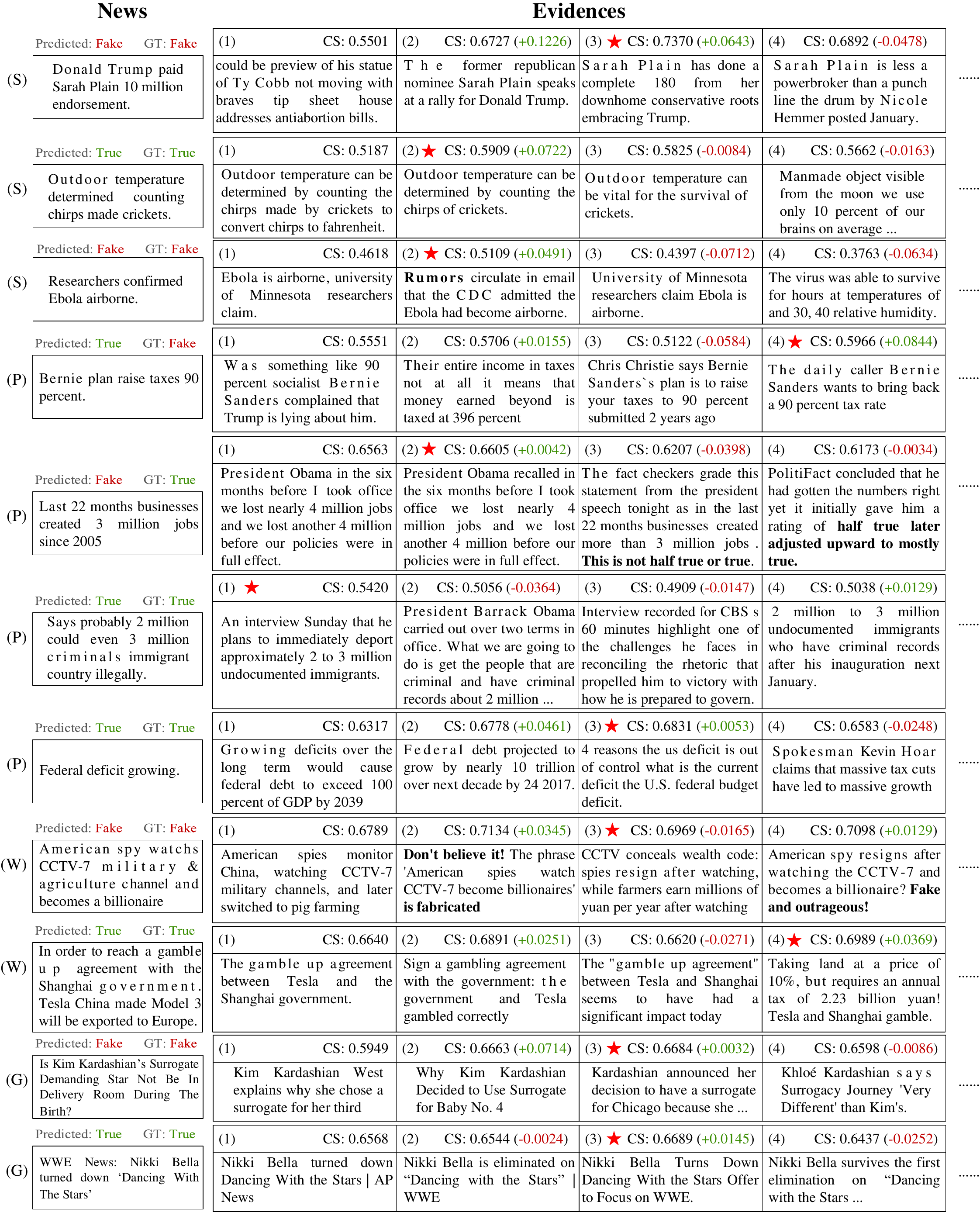}
  \caption{Case analysis on how the confidence score (upper right, denoted as CS) changes sequentially with the evidence at each time-step. Dataset: (W) Weibo21, (G) GossipCop, (S) Snopes, (P) PolitiFact. The time-steps having the highest confidence score (starred) both refer to the evidences which supports the correct prediction. 
  }
  \label{fig:caseanalysis}
\end{figure*}

Fig.~\ref{fig:caseanalysis} exhibits several cases to illustrate how the confidence score fluctuates sequentially with the input evidence at each time-step. 
The starred time-steps remark when SEE finishes the predictions.
We visually denote the explicit relationships between the evidence and predictions, with light yellow indicating support and light red signifying refutation or a lack of pertinent information. 
We study the implication of confidence score by reporting its variation between evidence, marked upper right for every time-step.
 
We see that those evidences usually cause a decrease in confidence score if it is of a different stance with the already-held one (marked in light red), which shows that SEE is aware of different stances of evidence pieces. 
Even though the confidence score gains or decreases over time given mutually-repelling materials, the magnitudes of the fluctuations also vary subjected to the specific context.
We further highlight words that provide information that is directly related to the veracity of news by bold text, e.g., ``fabricated'', ``rumors''. 
According to observation, evidence with such pristine information causes maximum variation among other evidence.
From the examples, we empirically derive that 1) the confidence score reflects the stubbornness of SEE holding a detection stance. 2) Adjacent pieces of evidence that hold consistent stances will raise the score while holding contradicting stances will decrease the score. And 3) the fluctuation amplitude of the confidence score is determined by SEE's learned knowledge about the evidence's usefulness.

\bibliography{mybibfile}